\documentclass[preprint2,tighten]{aastex62}
\usepackage{graphicx,bm}
\usepackage{rotating}

\newcommand{\lya}{\mbox{${\rm Ly}\alpha$}}

\providecommand{\kms}{\,\ensuremath{\rm{km\,s}^{-1}}}
\newcommand{\cmjj}{\mbox{${\rm cm^{-2}}$}}
\newcommand{\etal}{et al.}

\submitjournal{ApJL}

\shorttitle{}
\shortauthors{Chen et al.}

\begin{document}

\title{A Giant Intragroup Nebula Hosting a Damped \lya\ Absorber at
  $z=0.313$}

\correspondingauthor{Hsiao-Wen Chen}
\email{hchen@oddjob.uchicago.edu}

\author[0000-0001-8813-4182]{Hsiao-Wen Chen}
\affil{Department of Astronomy \& Astrophysics, The University of Chicago, 5640 S Ellis Ave., Chicago, IL 60637, USA}

\author[0000-0003-3244-0409]{Erin Boettcher}
\affiliation{Department of Astronomy \& Astrophysics, The University of Chicago, 5640 S Ellis Ave., Chicago, IL 60637, USA}

\author[0000-0001-9487-8583]{Sean D.\ Johnson}
\affiliation{Department of Astrophysics, Princeton University, Princeton, NJ, USA}
\affiliation{The Observatories of the Carnegie Institution for Science, 813 Santa Barbara Street, Pasadena, CA 91101, USA}
\affiliation{Hubble, Princeton--Carnegie fellow}

\author[0000-0001-7869-2551]{Fakhri S.\ Zahedy}
\affiliation{Department of Astronomy \& Astrophysics, The University of Chicago, 5640 S Ellis Ave., Chicago, IL 60637, USA}

\author{Gwen C.\ Rudie}
\affiliation{The Observatories of the Carnegie Institution for Science, 813 Santa Barbara Street, Pasadena, CA 91101, USA}

\author[0000-0001-5810-5225]{Kathy L.\ Cooksey}
\affiliation{Department of Physics and Astronomy, University of Hawai’i at Hilo, Hilo, HI 96720, USA}

\author{Michael Rauch}
\affiliation{The Observatories of the Carnegie Institution for Science, 813 Santa Barbara Street, Pasadena, CA 91101, USA}

\author[0000-0003-2083-5569]{John S.\ Mulchaey}
\affiliation{The Observatories of the Carnegie Institution for Science, 813 Santa Barbara Street, Pasadena, CA 91101, USA}



\begin{abstract}

This paper reports the discovery of spatially-extended line-emitting
nebula, reaching to $\approx 100$ physical kpc (pkpc) from a damped
\lya\ absorber (DLA) at $z_{\rm DLA}=0.313$ along the sightline toward
QSO PKS\,1127$-$145 ($z_{\rm QSO}=1.188$).  This DLA was known to be
associated with a galaxy group of dynamical mass $M_{\rm group}\sim
3\times 10^{12}\,{\rm M}_\odot$, but its physical origin remained
ambiguous.  New wide-field integral field observations
revealed a giant nebula
detected in [O\,\small{II}], H$\beta$, [O\,\small{III}], H$\alpha$, and
[N\,\small{II}] emission, with the line-emitting gas following closely
the motions of group galaxies.  One of the denser streams passes
directly in front of the QSO with kinematics consistent with the
absorption profiles recorded in the QSO echelle spectra.  The emission
morphology, kinematics, and line ratios of the nebula suggest that shocks
and turbulent mixing layers, produced as a result of stripped gaseous
streams moving at supersonic speed across the ambient hot medium,
contribute significantly to the ionization of the gas.  While the DLA
may not be associated with any specific detected member of the group, both
the kinematic and dust properties are consistent with the DLA
originating in streams of gas stripped from sub-$L_*$ group members at
$\lesssim 25$ pkpc from the QSO sightline.  This study demonstrates
that gas stripping in low-mass galaxy groups is effective in
releasing metal-enriched gas from star-forming regions, producing
absorption systems in QSO spectra, and that combining absorption and
emission-line observations provides an exciting new opportunity for
studying gas and galaxy co-evolution.


\end{abstract}




\section{Introduction} \label{sec:intro}

The diffuse circumgalactic and intergalactic gas beyond visible galaxy
disks contains $>90$\% of all baryons in the universe (e.g.,
Miralda-Escud\'e \etal\ 1996; Fukugita 2004), serving as a reservoir
of materials for sustaining the growth of galaxies while maintaining a
record of feedback from previous episodes of star formation and active
galactic nuclei (AGN) activity.  A comprehensive understanding of the
origin and evolution of galaxies relies on accurate characterizations
of this diffuse gas (e.g., Somerville \& Dav\'e 2015).  However,
direct imaging of the diffuse circumgalactic medium (CGM) around
distant galaxies has been challenging (e.g., Fynbo \etal\ 1999;
Christensen \etal\ 2006; Rauch \etal\ 2008, 2011, 2013; Steidel
\etal\ 2011; Krogager \etal\ 2017), because the gas density is
typically too low to radiate at sufficiently high intensities (e.g.,
Cantalupo \etal\ 2005; Kollmeier \etal\ 2010) and because the factor
of $(1+z)^4$ cosmological surface brightness dimming further
suppresses the signal.  On the other hand, quasar absorption
spectroscopy has provided a sensitive probe of tenuous gas based on
the absorption features imprinted in the quasar spectra (e.g., Rauch
1998; Wolfe \etal\ 2005; Chen 2017; Tumlinson \etal\ 2017).

Among different types of QSO absorption-line systems, damped
\lya\ absorbers (DLAs) with neutral hydrogen column density $N({\rm
  HI})\!\ge\!2\!\times\!10^{20}\,\cmjj$ are of particular significance
in understanding the complex interface between star formation and the
interstellar gas.  The high $N({\rm HI})$ ensures that the gas is
neutral which, together with the presence of heavy elements,
makes DLAs a promising
signpost of distant galaxies.  Studies of DLAs not only help
characterize the cosmic evolution of neutral gas with time (e.g.,
Neeleman \etal\ 2016), but also provide key insights into the
interplay between star formation in the interstellar
matter (ISM) of distant galaxies and their large-scale neutral gas properties (e.g., Wolfe \etal\ 2005).

Here we present a unique system, for which newly available
wide-field integral field spectroscopic (IFS) data enabled
detections of spatially extended line-emitting gas, connected
directly to a previously known DLA at $z_{\rm DLA}=0.313$.  It
demonstrates the power of combining emission and absorption
measurements, which provides a new and exciting opportunity for
studying gas and galaxy co-evolution (cf.\ Zhang \etal\ 2018).
Throughout the paper, we adopt a standard $\Lambda$ cosmology,
$\Omega_M$ = 0.3 and $\Omega_\Lambda$=0.7 with a Hubble constant
$H_{\rm 0} = 70\rm \,km\,s^{-1}\,Mpc^{-1}$.

\section{Observations and Data Descriptions} \label{sec:ions}

Here we briefly describe available imaging and spectroscopic data for
characterizing the DLA and its surrounding gas and galaxy properties.

\begin{figure}
  \begin{center}
\includegraphics[scale=0.48]{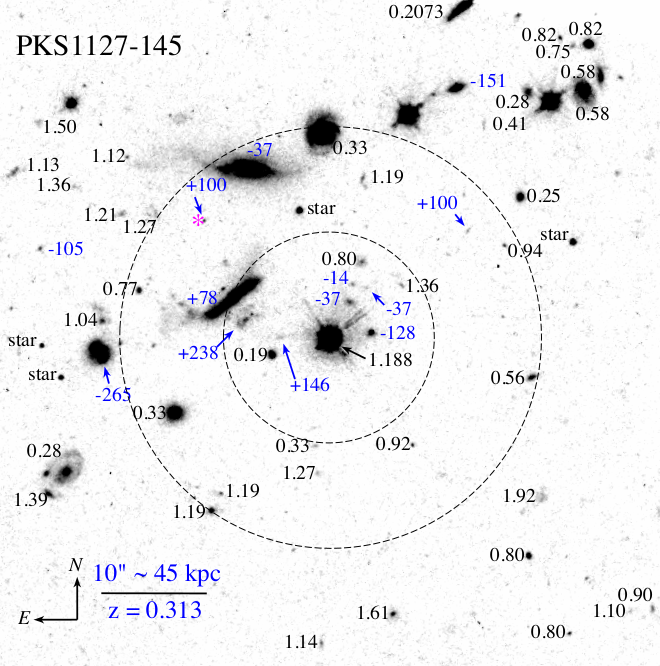}
\end{center}
\caption{Summary of the redshift survey enabled by MUSE
  data.  The image, obtained using {\it HST} WFPC2 and the F814W
  filter (PID$=$9173), covers a sky area of $\approx 1'\times 1'$
  centered at QSO PKS\,1127$-$145 at $z_{\rm QSO}=1.188$.  It reaches
  a 5-$\sigma$ limiting magnitude of $AB({\rm F814W})=27.4$ over a
  $0.5''$ diameter aperture.  The dashed circles mark the $10''$ and $20''$
  radii from the QSO sightline.  Spectroscopically identified objects
  from the MUSE data are marked with redshift measurements.  The
  redshifts of galaxies associated with the DLA at $z=0.313$ have been
  converted to line-of-sight velocity offsets from the fiducial
  redshift $z=0.31266$ (see Figure 2 below), which are shown in blue.
  The light-weighted
  center of the galaxy group, at ($+12.2''$, $+10.6''$) from the QSO, is marked by
  '*' in the image, corresponding to 74 pkpc in projected distance (see the main text for details).}
\end{figure}

\subsection{Galaxy imaging and spectroscopic data}

High spatial resolution optical images obtained using the Wide Field
and Planetary Camera 2 (WFPC2) and the F814W filter on board the {\it
  Hubble Space Telescope} ({\it HST}) were retrieved from the {\it
  HST} data archive (PID$=$9173; PI: Bechtold), 
and coadded to form a final combined image using custom
software.  Source detection was performed using the SExtractor program
(Bertin \& Arnouts 1996).  A portion of the combined image, covering
the $\approx 1'\times 1'$ area centered at the QSO, is presented in
Figure 1.

Wide-field IFS data of the field around PKS\,1127$-$145 were
obtained using the Multi-Unit Spectroscopic Explorer (MUSE; Bacon et
al.\ 2010).  MUSE observes a field of $1'\times 1'$ with a pixel scale
of $0.2''$ and a spectral resolution of ${\rm FWHM}\approx 120$
\kms\ at 7000 \AA.  The observations were obtained under 
PID$=$096.A-0303 (PI: P\'eroux), with a total integration of
8700 seconds and a mean seeing condition of ${\rm FWHM}\approx 0.6''$.
Pipeline-processed and flux-calibrated individual data cubes were
downloaded from ESO Phase 3 data archive.  The individual data cubes
were combined using custom software with optimal weights determined
based on the inverse variance of the sky and with the world
coordinates calibrated to match the {\it HST} WFPC2 image.  The
wavelength array was converted to vacuum.  Finally, a median sky
spectrum was formed using spaxels in blank areas and subtracted
from the final combined data cube to reduce sky residuals.

The combined MUSE data cube reaches a 1-$\sigma$ limiting sensitivity
over a $1''$ box of $1.7\times 10^{-19}\,{\rm erg}\,{\rm s}^{-1}\,{\rm
  cm}^{-2}\,{\rm \AA}^{-1}\,{\rm arcsec}^{-2}$ at 4900 \AA\ and 8600
\AA, and $8\times 10^{-20}\,{\rm erg}\,{\rm s}^{-1}\,{\rm
  cm}^{-2}\,{\rm \AA}^{-1}\,{\rm arcsec}^{-2}$ at 6600 \AA.  It
enables a highly complete spectroscopic survey of faint galaxies in
the QSO field, including those in the vicinity of the DLA at $z_{\rm
  DLA}=0.313$.  We extracted object spectra from the final MUSE data
cube using a circular aperture of two pixels ($0.4''$) in radius.
Redshift measurements were based on a $\chi^2$ fitting routine
described in Chen \& Mulchaey (2009) and Johnson \etal\ (2013).  The
best-fit redshift of each object 
was visually inspected for confirmation.  This exercise led to robust
redshift measurements
of 63 objects, 13 of which are 
found within $30''$ of the DLA. 
The redshift measurements are
presented in Figure 1.  To avoid crowding, redshifts of galaxies not
associated with the DLA are presented to two decimal places.  For the
galaxies associated with the DLA, the line-of-sight velocity offsets
are shown relative to $z=0.31266$ (see below).

\begin{figure*}
  \begin{center}
    \includegraphics[scale=0.6]{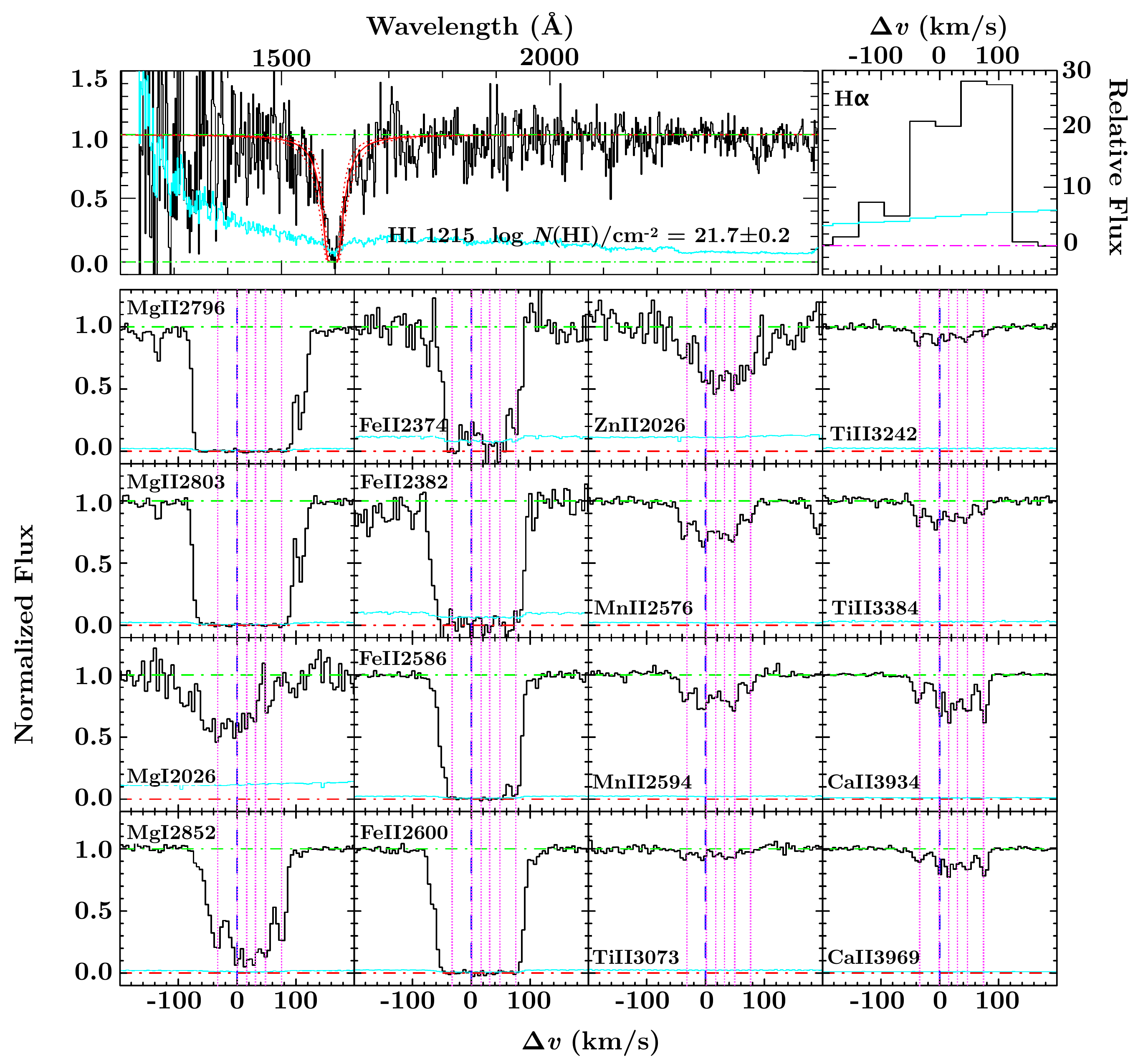}
    \end{center}
\caption{Comparison of gas kinematics in emission and in absorption.
  The top-left panel displays the continuum-normalized {\it HST} FOS
  spectrum in black, and the associated 1-$\sigma$ error spectrum in
  cyan.  The strong absorption at $\approx 1500$ \AA\ is the hydrogen
  \lya\ absorption feature with the best-fit Voigt profiles of
  $\log\,N({\rm H\,I})/{\rm cm}^{-2}=21.7\pm 0.2$ (shown in red; see
  also Rao \& Turnshek 2000).  The bottom panels display metal
  absorption lines associated with the DLA from UVES and STIS
  observations.  Zero velocity corresponds to $z = 0.31266$.   While strong
  transitions such as Mg\,II and Fe\,II are saturated, unsaturated
  transitions such as Mg\,I, Mn\,II, Ti\,II, and Ca\,II are well
  resolved into six individual components (vertical dotted lines; see also Guber \etal\ 2018).
  For comparison,
  the H$\alpha$ line from the line-emitting gas at
  $1.5''$ east of the QSO sightline is shown in the upper-right panel.
  While the emission spectrum does not have a sufficient spectral
  resolution for resolving small-scale motions, both the velocity
  centroid and width of the emission from the gaseous stream near the QSO sightline agree well
  with the mean and dispersion of the absorbing components.}
\end{figure*}

\subsection{Absorption spectra of the QSO}

Absorption spectra of PKS\,1127$-$145 were obtained using
the Faint Object Spectrograph (FOS) and
the low-resolution G160L grating on board {\it HST} (PID$=$6577; PI:
Rao), the Space Telescope Imaging Spectrograph (STIS) and the E230M
grating (PID$=$9173; PI: Bechtold), and the Ultraviolet and Visual
Echelle Spectrograph (UVES; D’Odorico et al.\ 2000) on the VLT-UT2
telescope under multiple programs (67.A-0567, 69.A-0371, 076.A-086).
Details regarding the STIS and UVES data processing are described in Cooksey \etal\ (2010) and Zahedy \etal\ (2017), respectively.

These ultraviolet and optical spectra enabled detailed studies of the
chemical and dust content of the DLA (e.g., Kanekar \etal\ 2014; Guber
\etal\ 2018).  In particular, the high-resolution echelle spectra
provide the wavelength coverage for observing a suite of heavy ions,
including Mg$^+$, Fe$^+$, Mn$^+$, Ti$^+$, and Ca$^+$.  A summary of
the absorption properties of the DLA is presented in Figure 2.
Metal absorption lines, such as Mn\,II, Ti\,II, and
Ca\,II, are resolved into at least six individual components.   The
strongest Mn\,II component occurs at $z_c=0.31266$, which is adopted as the
redshift zero point for subsequent discussions.  The remaining
components are detected at relative line-of-sight velocities, $\Delta\,v_c=-36$,
$+15$, $+27$, $+46$, and $+76$ \kms (vertical dotted lines in Figure
2).

\begin{table*}
\scriptsize
\centering
\caption{Galaxy properties in the vicinity of the DLA at $z=0.313$}
\label{table:sample}
\centering {
\begin{tabular}{lcrrrrrrrccr}
\hline \hline
                       &     & \multicolumn{1}{c}{$\Delta\,\alpha$} & \multicolumn{1}{c}{$\Delta\,\delta$} & \multicolumn{1}{c}{$\theta$} & \multicolumn{1}{c}{$d$}    & \multicolumn{1}{c}{$\Delta\,v_g^a$} & \multicolumn{1}{c}{$AB({\rm F814W})$} & \multicolumn{1}{c}{$M_r^b$} & \multicolumn{1}{c}{$L_r$} &  & \\
\multicolumn{1}{c}{ID} & $z$ & \multicolumn{1}{c}{($''$)}           & \multicolumn{1}{c}{($''$)}           & \multicolumn{1}{c}{($''$)}   & \multicolumn{1}{c}{(pkpc)} & \multicolumn{1}{c}{(\kms)}      & \multicolumn{1}{c}{(mag)}             & \multicolumn{1}{c}{(mag)} & \multicolumn{1}{c}{($L_*$)}& \multicolumn{1}{c}{Type$^c$}& \multicolumn{1}{c}{crossID$^d$} \\
\multicolumn{1}{c}{(1)} & \multicolumn{1}{c}{(2)}  & \multicolumn{1}{c}{(3)}  & \multicolumn{1}{c}{(4)} & \multicolumn{1}{c}{(5)} & \multicolumn{1}{c}{(6)}& \multicolumn{1}{c}{(7)}& \multicolumn{1}{c}{(8)}& \multicolumn{1}{c}{(9)}& \multicolumn{1}{c}{(10)}& \multicolumn{1}{c}{(11)} & \multicolumn{1}{c}{(12)}\\
\hline
J113006.80$-$144926.86 &  0.3121 &  $-$3.8 &  $+$0.4 &   3.8 &  17.4 & $-$128 & 22.11 & $-18.66$ & 0.09 & em &  G1  \\ 
J113006.92$-$144923.96 &  0.3125 &  $-$2.0 &  $+$3.3 &   3.9 &  17.9 &  $-$37 & 22.19 & $-18.58$ & 0.08 & abs &   \\ 
J113007.38$-$144927.06 &  0.3133 &  $+$4.6 &  $+$0.2 &   4.6 &  21.3 & $+$146 & 24.56 & $-16.21$ & 0.01 & em & G16  \\ 
J113006.89$-$144922.36 &  0.3126 &  $-$2.5 &  $+$4.9 &   5.5 &  25.3 &  $-$14 & 23.88 & $-16.89$ & 0.02 & em &   \\ 
J113006.82$-$144922.63 &  0.3125 &  $-$3.5 &  $+$4.7 &   5.8 &  26.7 &  $-$37 & 24.67 & $-16.10$ & 0.01 & em & G18  \\ 
J113007.61$-$144925.49 &  0.3137 &  $+$8.0 &  $+$1.8 &   8.2 &  37.5 & $+$238 & 21.79 & $-18.98$ & 0.12 & em & G17  \\ 
J113007.68$-$144923.29 &  0.3130 &  $+$9.0 &  $+$4.0 &   9.8 &  45.1 &  $+$78 & 19.24 & $-21.53$ & 1.24 & em &  G2  \\ 
J113007.86$-$144916.36 &  0.3131 & $+$11.6 & $+$10.9 &  15.9 &  73.1 & $+$100 & 25.18 & $-15.59$ & 0.01 & em &   \\
J113006.16$-$144917.16 &  0.3131 & $-$13.1 & $+$10.1 &  16.5 &  75.8 & $+$100 & 25.15 & $-15.62$ & 0.01 & em &   \\
J113007.58$-$144911.01 &  0.3125 &  $+$7.5 & $+$16.3 &  17.9 &  82.2 &  $-$37 & 18.95 & $-21.82$ & 1.62 & LINER  &  G4  \\ 
J113008.55$-$144928.26 &  0.3115 & $+$21.6 &  $-$1.0 &  21.6 &  98.8 & $-$265 & 20.20 & $-20.57$ & 0.51 & em &  G6  \\ 
J113006.23$-$144903.80 &  0.3120 & $-$12.0 & $+$23.5 &  26.4 & 120.8 & $-$151 & 21.32 & $-19.45$ & 0.18 & em & G19  \\ 
J113008.92$-$144918.30 &  0.3122 & $+$27.0 &  $+$9.0 &  28.4 & 130.1 & $-$105 & 24.51 & $-16.26$ & 0.01 & em & G21  \\ 
J113010.32$-$144904.34 &  0.3124$^e$ & $+$47.3 & $+$23.0 &  52.5 & 240.6 &  $-$59 & 20.42 & $-20.35$ & 0.42 & em & G14  \\ 
\hline
\multicolumn{10}{l}{$^\mathrm{a}$$\Delta\,v_g=0$ corresponds to $z=0.31266$, the strongest absorbing component in Figure 2.} \\
\multicolumn{10}{l}{$^\mathrm{b}$We adopt a characteristic rest-frame absolute $r$-band magnitude of $M_{r_*}=-21.3$ from Cool \etal\ (2012).} \\
\multicolumn{10}{l}{$^\mathrm{c}$Spectral type of the galaxy: emission-line ('em'), absorption ('abs'), or LINER dominated.} \\
\multicolumn{10}{l}{$^\mathrm{d}$Galaxy ID's from Kacprzak \etal\ (2010) and P\'eroux \etal\ (2019).} \\
\multicolumn{10}{l}{$^\mathrm{e}$This galaxy occurs outside the MUSE field of view.  Redshift is taken from Kacprzak \etal\ (2010).} \\
\end{tabular}
}
\end{table*}

\section{The Galactic Environment of the DLA at ${z_{\rm DLA}=0.313}$}

The MUSE observations described in \S\ 2.1 have yielded a total of 13
galaxies in the vicinity of the DLA at $z_{\rm DLA}=0.313$ toward
PKS\,1127$-$145 (cf.\ P\'eroux \etal\ 2019), with redshifts ranging
from $z=0.3115$ to $z=0.3137$ and physical projected distances from
$d=17.4$ pkpc to $d=130$ pkpc.  Including the galaxy found at
$z=0.3124$ and $d=241$ kpc by Kacprzak \etal\ (2010), we establish a
total of 14 galaxies associated with the DLA, four of which are new
from this work (cf.\ Kacprzak \etal\ 2010; P\'eroux \etal\ 2019).
Table 1 presents a complete list of galaxies spectroscopically
identified in the vicinity of the DLA.  In columns (1) through (11),
we present the galaxy ID based on their J2000 coordinates, best-fit
redshift, angular offsets in right ascension and declination
($\Delta\,\alpha$, $\Delta\,\delta$) and angular distance ($\theta$)
of the galaxy from the background QSO, the projected distance ($d$) in
physical units from the QSO sightline, the line-of-sight velocity
offset from the fiducial redshift $z=0.31266$ ($\Delta\,v_g$), the
observed $AB({\rm F814W})$ magnitude, the corresponding rest-frame
$r$-band absolute magnitude ($M_r$), the intrinsic luminosity in units
of $L_*$, and cross-reference ID from Kacprzak \etal\ (2010) and
P\'eroux \etal\ (2019)\footnote{Note that G20 was identified by
  P\'eroux \etal\ 2019 as a group member based on the absorption
  features, but we find it to be a star.}.

The galaxy group in the vicinity of the DLA spans a range in the
optical brightness, from $AB({\rm F814W})\approx 19$ to $AB({\rm
  F814W})\approx 25.2$ mag.  At $z=0.313$, the observed F814W band
corresponds to the rest-frame $r$-band.  The observed brightnesses
therefore translate directly to the rest-frame $r$-band luminosities.
Adopting the characteristic rest-frame absolute $r$-band magnitude of
$M_{r_*}=-21.3$ for blue galaxies from Cool \etal\ (2012), the
corresponding rest-frame luminosities range from $\lesssim\,0.01\,L_*$
to $\approx\,1.6\,L_*$ (see also Kacprzak \etal\ 2010; P\'eroux
\etal\ 2019).  In addition, the observed line-of-sight velocity
offsets of the group members relative to the strongest absorption
component at $z=0.31266$ range from $\Delta\,v_g=-265$ to
$\Delta\,v_g=+238$ \kms.

We also compute the light-weighted center of the group at ($+12.2''$,
$+10.6''$) from the QSO (marked by '*' in Figure 1), with a
corresponding projected distance of $d_{\rm group}=74$ pkpc from the
DLA.  The observed line-of-sight velocity dispersion of
$\sigma_v=128$ \kms\ among the 14 group members indicates a dynamical
group mass of $M_{\rm group}\sim 3\times 10^{12}\,{\rm M}_\odot$.

The presence of a galaxy group around the DLA complicates common theories
of the physical origin of the high-column density gas.  With a mean
gas metallicity $5\times$ lower in the DLA than in the ISM of luminous
group members, previously favored scenarios include low-luminosity
dwarf satellites (e.g., York \etal\ 1986), tidal debris in the group
environment (e.g., Kacprzak \etal\ 2010), and diffuse intragroup
medium (e.g., P\'eroux \etal\ 2019).
  
\begin{figure*}
  \begin{center}
    \includegraphics[scale=0.9]{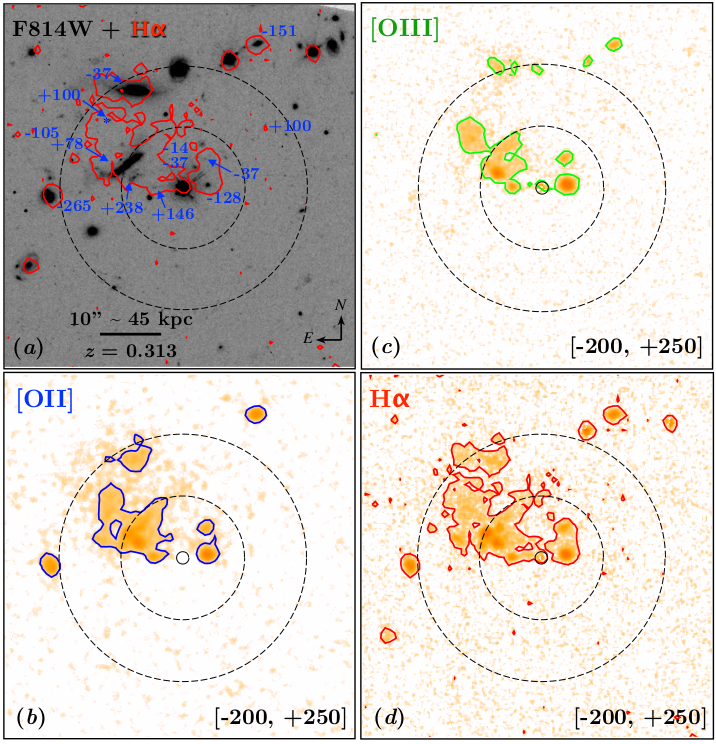}
    \end{center}
  \caption{The spatial distribution of line-emitting gas relative to
    the stellar continuum.  Panel ({\it a}) shows the {\it HST} F814W
    image of the field around PKS\,1127$-$145 with spectroscopically
    identified galaxy group members marked by their line-of-sight
    velocity offsets.  The continuum subtracted line-flux maps of
    [O\small{II}]$\lambda\lambda\,3727, 3729$,
    [O\,\small{III}]$\lambda\,5007$, and H$\alpha$ integrated over a
    velocity window from $\Delta\,v=-200$ \kms\ to $\Delta\,v=+250$
    \kms\ are displayed in panels ({\it b}), ({\it c}), and ({\it d})
    with the contours representing the constant surface brightness of
    $2.5\times 10^{-17}\,{\rm erg}\,{\rm s}^{-1}\,{\rm cm}^{-2}\,{\rm
      arcsec}^{-2}$.  The H$\alpha$ contour from panel ({\it d}) is
    superimposed on the F814W image in panel ({\it a}) to highlight
    that the line-emitting gas extends much beyond the stellar disks.}
\end{figure*}

However, in addition to individual galaxies associated with DLA, the
MUSE data also uncovered spatially extended line-emitting gas,
detected in H$\alpha$, H$\beta$, [O\,\small{III}], and
[O\,\small{II}], across the galaxy group.  The emitting morphology is
consistent with the gas being stripped from members of the galaxy
group, similar to what was found in other group environments (e.g.,
Epinat \etal\ 2018; Johnson \etal\ 2018).  One of the denser streams
passes directly in front of the QSO with kinematics consistent with
the absorption profiles revealed in the QSO echelle spectrum.  While
the MUSE data do not have the spectral resolution necessary to resolve
relative motions under $\approx 100$ \kms, both the velocity centroid and
width of the lines near the QSO sightline agree well with the
mean and dispersion of the absorbing components (top-right panel
in Figure 2).  The imaging panels in Figure 3 compare the spatial
distributions of the stellar continuum revealed in the {\it HST} F814W
image and the line-emitting nebula from the MUSE data.  Continuum
subtracted line-flux maps of [O\small{II}]$\lambda\lambda\,3727,
3729$, [O\,\small{III}]$\lambda\,5007$, and H$\alpha$ integrated over
a velocity window from $\Delta\,v=-200$ \kms\ to $\Delta\,v=+250$
\kms\ are displayed.  The line-emitting gas, detected in multiple
transitions, is seen to extend much beyond the stellar disks, reaching
to $\approx 100$ pkpc from the QSO sightline.

In addition, this extended gas is seen to follow a coherent motion
that closely traces the motion of the galaxies in the group, with the
intensity peaks moving from $\approx -200$ \kms\ west to
$\approx +250$ \kms\ east of the QSO ({\it left} panels of Figure 4).
Furthermore, the gas exhibits a broad range in velocity dispersion.
The observed large dispersion in [O\,\small{III}], in
particular, indicates the presence of turbulent, multiphase intragroup
gas (cf.\ Johnson
\etal\ 2018; see below).  We measure a total integrated line flux in
H$\alpha$ of $f({\rm H\alpha})=(1.38\pm 0.01)\times 10^{-15}\,{\rm
  erg}\,{\rm s}^{-1}\,{\rm cm}^{-2}$ and in [O\,\small{III}] of
$f({\rm [OIII]})=(8.06\pm 0.05)\times 10^{-16}\,{\rm erg}\,{\rm
  s}^{-1}\,{\rm cm}^{-2}$.


\begin{sidewaysfigure*}
  \centering
  \vspace{+3.5in}
  \includegraphics[width=0.9\textwidth]{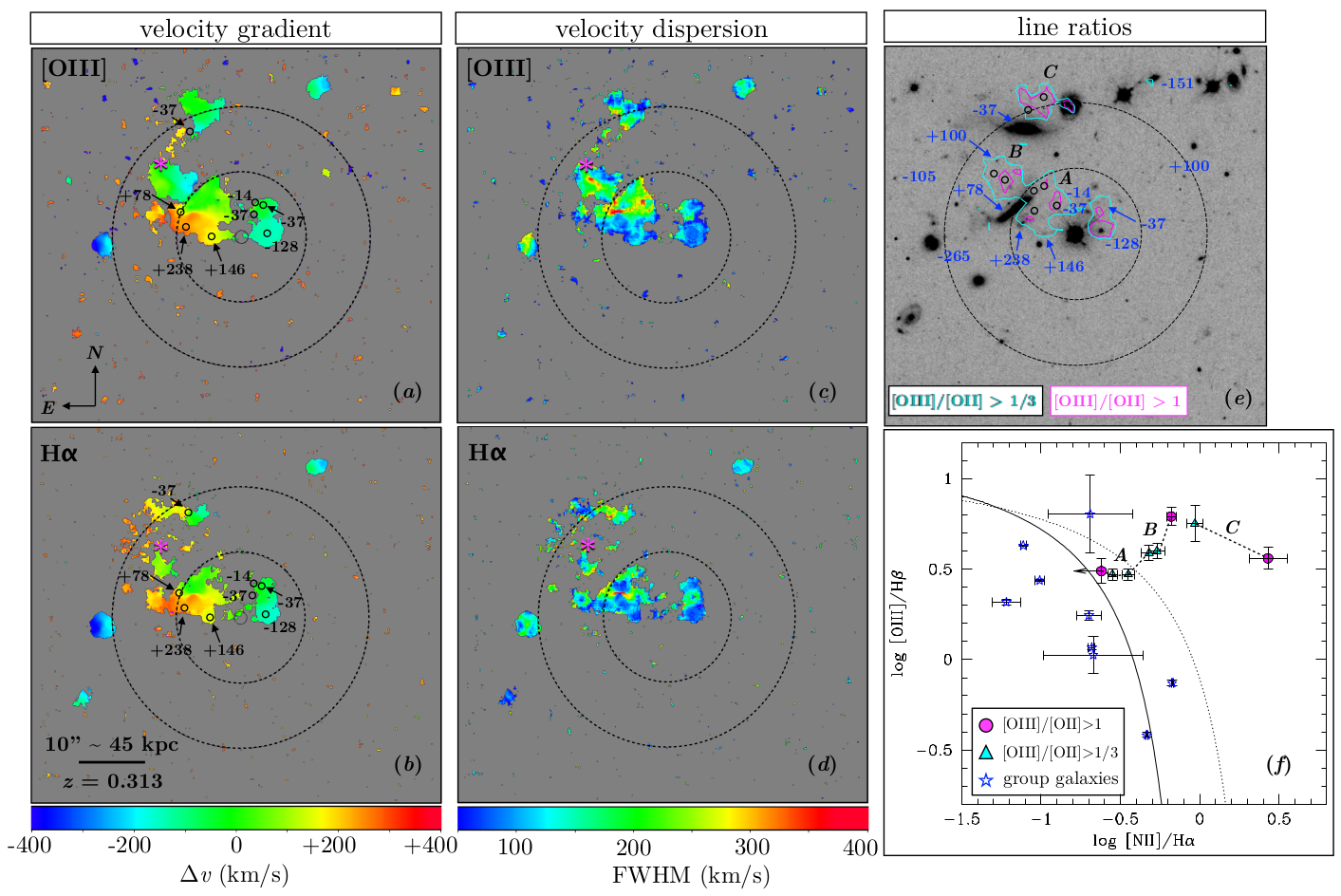}
\caption{({\it Left} column) Velocity gradient ($\Delta\,v$) of the
  [O\small{III}]$\lambda\,5007$ (panel {\it a}) and H$\alpha$ (panel
  {\it b}) emitting gas, with the corresponding best-fit velocity
  dispersion (FWHM in \kms), after removing the instrument line width,
  displayed in panels ({\it c}) and ({\it d}), respectively, in the
  {\it middle} column.  Zero velocity corresponds to $z=0.31266$.  The
  positions of the galaxies that coincide with the nebula are marked
  by open circles along with their line-of-sight velocity offsets,
  showing that the line-emitting gas following closely the motions of
  group galaxies.
  Panel ({\it e}) highlights the locations where 
  [O\small{III}]/[O\small{II}]$>1/3$ (cyan contours) and
  $>1$ (magenta contours), relative to group  galaxies.
  Panel ({\it f}) shows the observed
  [O\small{III}]$\lambda\,5007$/H$\beta$ versus
  [N\small{II}]$\lambda\,6585$/H$\alpha$ ratios for three regions,
  $A$, $B$, and $C$, in Panel ({\it e}), at angular distances of
  $<10''$, between $10''$ and $20''$, and $>20''$, respectively.  The
  measurements are made based on integrated line fluxes over a
  $1''$-diameter aperture, and the positions of the apertures are
  marked by open circles in Panel ({\it e}).  For comparison,
  the line ratios observed in the ISM of the group galaxies are also
  included.  Error bars represent the 1-$\sigma$ measurement
  uncertainties.  The solid and dotted curves are from Kauffmann
  \etal\ (2003) and Kewley \etal\ (2001), respectively, separating
  star-forming regions (lower-left) from active galaxies
  (upper-right).}
\end{sidewaysfigure*}

\section{Discussion} \label{sec:discussion}

Newly available wide-field IFS observations of the field around
PKS\,1127$-$145 have provided a wealth of information concerning the
galactic environment of the DLA at $z_{\rm DLA}=0.313$.  While this
DLA is previously known to be associated with a galaxy group (e.g.,
Bergeron \& Boiss\'e 1991, Chen \& Lanzetta 2003; Kacprzak et
al.\ 2010; P\'eroux et al.\ 2019),
the IFS data have uncovered spatially extended, line-emitting nebula
connecting between members of the galaxy group.  One of the denser
streams passes directly in front of the QSO with kinematics consistent
with the absorption profiles revealed in the QSO echelle spectrum,
establishing a direct connection between the DLA and the giant nebula.

The MUSE data also allow us to examine the physical condition of the gas based
on comparisons of multiple transitions.  First, we examine the gas
density based on the observed [O\small{II}] doublet ratio and the
H$\alpha$ surface brightness (e.g., Osterbrock \& Ferland 2006).  The
[O\small{II}]$\lambda\,3729$ line is found to be comparable or
stronger than the $\lambda\,3727$ member across the line-emitting
nebulae, constraining the electron density at $n_e \lesssim 300\,{\rm
  cm}^{-3}$.  At the same time, the observed H$\alpha$ surface
brightness is related to the gas density following,
\begin{equation}
{\rm SB}_{\rm H\alpha}\approx 1.7\times 10^{-15}\,C\,\frac{\langle n_e\rangle^2}{(1+z)^4}\,\frac{l_{\rm neb}}{{\rm kpc}}\,{\rm erg}\,{s}^{-1}\,{\rm cm}^{-2}\,{\rm arcsec}^{-2},
\end{equation}
where $C\equiv\langle n_e^2\rangle/\langle n_e\rangle^2$ represents
the clumping factor of the line-emitting gas, $\langle n_e\rangle$
represents the mean gas density, and $l_{\rm neb}$ represents the
depth of the line-emitting region in units of kpc.  Adopting a
characteristic surface brightness of $2.5\times 10^{-17}\,{\rm
  erg}\,{\rm s}^{-1}\,{\rm cm}^{-2}\,{\rm arcsec}^{-2}$ from Figure 3
and assuming $l_{\rm neb}=1$ kpc lead to a crude estimate of gas
density of $\langle\,n_e\,\rangle=0.2\,(0.06)\,{\rm cm}^{-3}$ for
$C=1\,(10)$.

Next, we examine the ionization state of the gas by measuring the
[O\small{III}]$\lambda\,5007$/[O\small{II}]$\lambda\lambda\,3727,
3729$ ratio in the nebula.  Panel ({\it e}) of Figure 4 shows the
contours of [O\small{III}]/[O\small{II}]$=1/3$ (cyan) and 1 (magenta).
These contours identify the locations of highly-ionized gas both
within 45 pkpc of the DLA and next to the two super-$L_*$ spirals in
the group with flared/warped disk morphologies.  Because the observed
[O\small{III}]/[O\small{II}] ratio depends on both the ionization
radiation and gas metallicity (e.g., Kewley \& Dopita 2002), we also
place this gas on the BPT diagram (Baldwin \etal\ 1981) to examine how
the observed [O\small{III}]$\lambda\,5007$/H$\beta$ compares with
[N\small{II}]$\lambda\,6585$/H$\alpha$
(e.g., Kewley
\etal\ 2013).

Panel ({\it f}) of Figure 4 shows line ratios at different locations
in the nebula, in comparison to the ISM of the group
galaxies\footnote{Note that J113007.58$-$144911.01 (G4 in P\'eroux
  \etal\ 2019) at 82 pkpc exhibits a LINER-like [NII]/H$\alpha$ but is
  not detected in H$\beta$ or [O\small{III}], preventing its inclusion
  in the panel.}.
Error bars show the 1-$\sigma$ measurement uncertainties.  While the
galaxies appear to be typical of distant star-forming galaxies on the
BPT diagram, the observed line ratios of high
[O\small{III}]/[O\small{II}] regions appear in the parameter space
traditionally occupied by active galaxies, indicating the need of a
hard ionizing spectrum to account for the ionization condition of the
gas (e.g., Kewley \etal\ 2013).  Specifically, the nebula at $>10''$
($>45$ pkpc) from the QSO sightline (regions $B$ and $C$ in Panel {\it e}
of Figure 4) appears to be under a more extreme ionizing condition
than the gas closer to the QSO sightline.


Possible sources for a hard radiation field include AGN, turbulent
mixing layers (e.g., Slavin \etal\ 1993; Miller \& Veilleux 2003), and
shocks (e.g., Dopita \& Sutherland 1995; Rich \etal\ 2011).  We have
searched for signatures of AGN in the group galaxies by examining
their optical spectra and WISE photometry (e.g., Wright \etal\ 2010;
Mateos \etal\ 2013), and found none.  The spatial variation of the
ionization state of the gas is also inconsistent with fossil ionized
regions from a previous AGN outburst.  On the other hand, the presence
of extended stellar streams from the two super-$L_*$ disk galaxies in
the {\it HST} image are indicative of ongoing violent galaxy
interactions between the group members.  For a dynamical mass of
$M_{\rm group}\sim 3\times 10^{12}\,M_{\odot}$, the temperature of the
intragroup medium would be $T\sim 2\times 10^6$ K, if present, and
the sound speed of the hot halo would be $c_s\sim 220$ \kms.  As
galaxies move through this hot intragroup medium, ram-pressure
stripping, in addition to tidal interactions between galaxy group
members, is expected to be effective in removing the ISM (e.g., Gunn
\& Gott 1972; Roediger \& Br\"uggen 2007) and shocks and/or turbulent
mixing layers should form as stripped gaseous streams travel through
the hot gas at supersonic speed.  Both the observed line widths seen
in [O\,\small{III}] and H$\alpha$ (panels {\it c} and {\it d}) and
emission morphologies are consistent with the expectation that
the line-emitting gas originates in shocks or turbulent mixing layers.
Ram-pressure and tidal stripping are most effective in removing gas in
the outskirts of galaxies (e.g., Roediger \& Br\"uggen 2007).  Coupled
with metallicity gradients commonly seen in galaxy disks (e.g.,
Bresolin 2017), this naturally explains the low metallicity in the
DLA.

In summary, available MUSE data have revealed a complex, multiphase
intragroup medium hosting a DLA at $z_{\rm DLA}=0.313$.  The DLA is
not associated with any specific detected member of the group.  Both
the kinematic and dust properties are consistent with the absorber
originating in streams of stripped gas from sub-$L_*$ group members at
$d\lesssim 25$ pkpc from the QSO sightline as a result of violent
interactions in the group environment.  In contrast, the nebula at
$d>45$ pkpc likely originates in the two super-$L_*$ spirals; and its
kinematics, line ratios, and morphologies are consistent with being
ionized by shocks or turbulent mixing layers.  This system
demonstrates that metal-enriched gas can be released from star-forming
regions into the intragroup medium through effective gas stripping
processes in distant low-mass groups, where it produces absorption
systems in QSO spectra.

\acknowledgments We thank an anonymous referee, C\'eline P\'eroux, and
Don York for helpful comments on an earlier draft.  HWC, EB, and FSZ
acknowledge partial support from NSF AST-1715692 and HST-GO-15163.01A
grants.  SDJ acknowledges support by NASA through a Hubble Fellowship
grant HST--HF2--51375.001--A.  KLC acknowledges support from NSF grant
AST-1615296.

%

\vspace{5mm}


\software{
          SExtractor (Bertin \& Arnout 1996)
          }

\end{document}